\documentclass[twocol]{ametsoc}

\usepackage{url}
\usepackage[pdftex]{epsfig}

\usepackage{color}
\newcommand{\remove}[1]{ }

\journal{bams}

\bibpunct{(}{)}{;}{a}{}{,}

\title{Evaluation, Tuning and Interpretation of Neural Networks for Meteorological Applications}

    \authors{Imme Ebert-Uphoff\correspondingauthor{Imme Ebert-Uphoff, 
     Cooperative Institute for Research in the Atmosphere, Colorado State University, Fort Collins, CO. USA.}
     \thanks{Electrical and Computer Engineering, Colorado State University, Fort Collins, CO. USA; and Cooperative Institute for Research in the Atmosphere, Colorado State University, Fort Collins, CO. USA}
 and Kyle Hilburn\thanks{Cooperative Institute for Research in the Atmosphere, Colorado State University, Fort Collins, CO. USA.}}

    \affiliation{}

\email{iebert@colostate.edu}

\abstract{
Neural networks have opened up many new opportunities to utilize remotely sensed images in meteorology. Common applications include image classification, e.g., to determine whether an image contains a tropical cyclone, and image translation, e.g., to emulate radar imagery for satellites that only have passive channels.
However, there are yet many open questions regarding the use of neural networks in meteorology, such as best practices for evaluation, tuning and interpretation.  This article highlights several strategies and practical considerations for neural network development that have not yet received much attention in the meteorological community, such as the concept of effective receptive fields, underutilized meteorological performance measures, and methods for NN interpretation, such as synthetic experiments and layer-wise relevance propagation.  We also consider the process of neural network interpretation as a whole, recognizing it as an iterative scientist-driven discovery process, and breaking it down into individual steps that researchers can take.  Finally, while most work on neural network interpretation in meteorology has so far focused on networks for image classification tasks, we expand the focus to also include networks for image translation.
}

\begin{document}

\maketitle


{\bf Capsule:} This article discusses strategies for the development of neural networks for meteorological applications. Topics include evaluation, tuning and interpretation of neural networks, with special emphasis on networks for satellite imagery.

\section{Introduction}

Neural networks (NNs) are increasingly emerging as useful tools for meteorological applications
\citep{boukabara2019leveraging,reichstein2019deep,lee2018machine}.
However, because of their novelty many questions are yet to be answered for their use, such as
\begin{itemize}
\item 
    {\bf Evaluation and Tuninng:} 
    Which performance measures are most useful to evaluate and tune the neural network in a meteorological context?
\item
    {\bf Interpretability:} If the NN is performing well, {\it how} does it generate  meteorologically meaningful results?  Can we discover the strategies it uses?
\end{itemize}
We consider both types of questions here, building on the work of others, most notably of \cite{McGovern2019-transparent}, 
\cite{toms2019physically}, 
\cite{Gagne2019-interpretable} and \cite{Gagne2015-CPD}. 

Developing a better understanding of NN models can benefit meteorological applications in the following ways:
i) identify and eliminate potential failure modes, and, more generally, identify ways to improve the model's overall performance;  
ii) improve trust in the model; and 
iii) gain scientific insights that might lead to the development of simpler, more transparent approaches for such tasks; and iv) even to learn new physics \citep{toms2019physically,barnes2019viewing}.

In this paper we seek to extend existing work by the following contributions:
\begin{itemize}
\item 
    Breaking down the process of NN evaluation and interpretation for meteorological applications into individual steps and enumerating several methods for each step;
\item
    Expanding the discussion to image-to-image translation networks, which have received little attention in the context of interpretation;
\item
    Highlighting two important NN concepts that have received little attention in meteorology, namely receptive fields and layer-wise relevance propagation.
\end{itemize}

A key lesson we learned in the course of this research was that 
{\it gaining insights into the neural network consisted of an iterative, scientist-driven discovery process, driven by old fashioned methods of experimental design, and hypothesis generation and testing, with NN visualization tools simply providing additional tools to assist this process.}

The discussion in this article is organized as follows.
Section \ref{architecture_sec} discusses NN architectures, 
Section \ref{performance_sec} highlights strategies for NN evaluation and tuning, 
Section \ref{interpretation_sec} presents NN interpretation via targeted experiments and 
Section \ref{ANN_visualization_tools} discusses selected NN visualization methods.
Section \ref{conclusions_sec} presents conclusions.


\section{NN Architectures for Working with Images in Meteorological Applications}
\label{architecture_sec}

\begin{figure*}
    \begin{center}
    \epsfxsize=0.80\hsize 
    \epsfbox{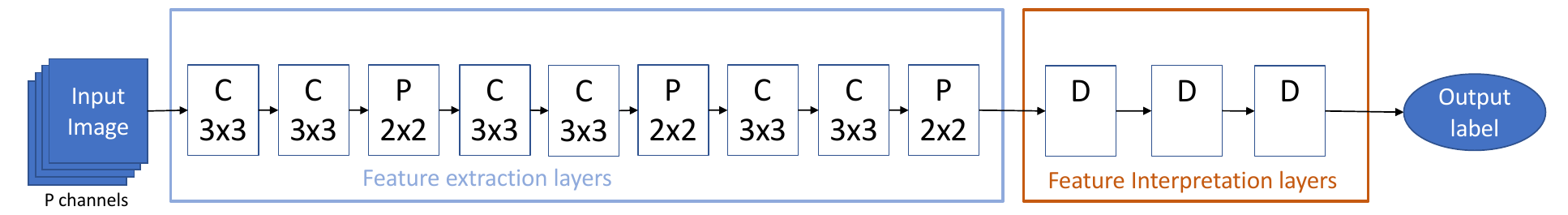}
    
    {\small (a) Sample CNN architecture for Classification}
    \vspace*{0.5cm}

    \epsfxsize=0.80\hsize 
    \epsfbox{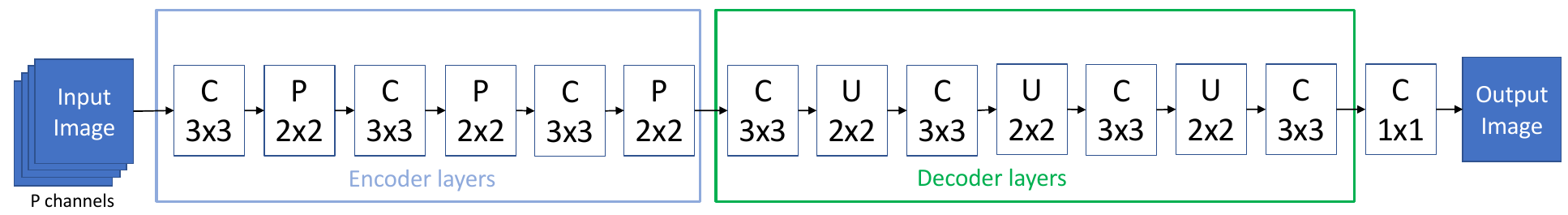}
    
    {\small (b) Sample Encoder-Decoder architecture for Image Translation}
    \vspace*{0.5cm}

    \epsfxsize=0.80\hsize 
    \epsfbox{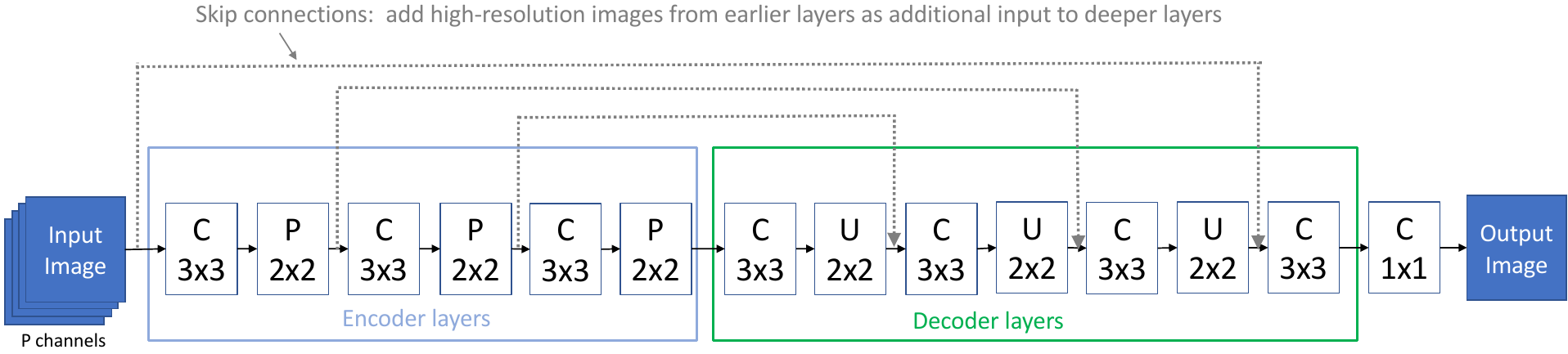}
    
    {\small (c) Sample U-net architecture for Image Translation}
    
    \end{center}
    
    \caption{Samples of three common NN architecture types for (a) classification and (b,c) image-to-image translation tasks. Here 'C' indicates a convolution layer, 'P' a pooling layer, 'U' an upsampling layer, and 'D' a fully-connected (aka dense) layer.  Sample filter dimensions are provided where relevant.}
    \label{architecture_fig}
\end{figure*}

Neural networks have opened up many new avenues to ingest and utilize images in meteorology, e.g., to identify specific patterns in an image (image classification) or to transform information in an image into a different representation (image-to-image translation).
Fig.\ \ref{architecture_fig} illustrates some common NN architectures for such tasks, 
%
%
which use four different layer types, convolution (C), max pooling (P), upsampling (U) and fully connected, aka dense, (D) layers.  
All three sample architectures shown in Fig.\ \ref{architecture_fig} take as input an image, typically consisting of several channels.  The classification model (Fig.\ \ref{architecture_fig}(a)) generates as output the most likely image label out of a number of predefined labels (aka classes).  Such networks can be used for example 
to estimate tropical cyclone intensity \citep{wimmers2019using},
identify cyclones \citep{bonfanti2018machine},
detect large hail \citep{Gagne2019-interpretable},
detect synoptic-scale fronts \citep{lagerquist2019deep},
anticipate tornadoes, and predict precipitation to occur in form of rain, freezing rain, snow, or ice \citep{McGovern2019-transparent}.

In contrast the image translation models (Fig.\ \ref{architecture_fig}(b,c)) generate as output an image, typically of the same dimension (but not necessarily the same number of channels) as the input image.  
Such image translation models provide a versatile tool to enhance remote sensing images \citep{tsagkatakis2019survey} and have been used to detect changes in satellite imagery \citep{peng2019end},
to emulate radar imagery for satellites that only have passive channels \citep{hilburn2020development}, and to identify all locations in a satellite image corresponding to convection \citep{yoonjin2020}.

All three architectures start with a sequence of convolution and pooling layers (blue box).
This sequence typically consists of repeating patterns of the form 'CP', 'CCP', etc., with the repeating pattern and number of repetitions depending on the application.  The purpose of this sequence is identical in all three models, namely to extract spatial patterns of increasing size from the input image.  Since convolution filters are limited in size (most efficient filter size tends to be (3x3)), the pooling layers serve the purpose of reducing resolution at their output, thus allowing subsequent convolution filters to act on pixels that represent larger and larger areas of the input space, and thus detecting larger patterns.  The increase in size of the detected patterns is studied in more detail in the receptive field subsection below.

In the classification model (Fig.\ \ref{architecture_fig}(a)) the feature extraction layers (blue box) first extract spatial patterns, aka features. After that the feature interpretation layers (red box) interpret the presence of the extracted features to generate a suitable output label, i.e.\ to determine the class the image belongs to.  
In contrast in the image translation models (Fig.\ \ref{architecture_fig}(b,c)) the blue box is followed by a sequence of convolution and upsampling layers (green box) which increases image resolution to restore original image size, while also unfolding the image typically into a different representation, i.e.\ translating the detected patterns into a different, learned representation of those patterns to generate the output image. (As an analogy, one may think of the task of translating from one language to another.  Given a sequence of words in one language, the encoder first extracts patterns (meaning) from the sequence, then the decoder expands those patterns in the other language.)
For images, because of the information lost during the downsampling process, skip layers are often added that allow high-resolution information to be added to the later layers as supplemental channels.  This type of architecture is called a U-net and is shown in Fig.\ \ref{architecture_fig}(c).   

A more advanced NN architecture for image translation are Generative Adversarial Networks (GANs), which are known to produce satellite images that look particularly realistic 
\citep{xu2018satellite}.
Probably the most popular type of GAN for satellite images is 
Pix2Pix \citep{isola2017image} which is a general-purpose image translation tool.
\cite{kim2019nighttime} use pix2pix to generate nighttime reflectance imagery from visible satellite bands. 
While the details of GANs are beyond the scope of this paper, Pix2Pix typically uses either an encoder-decoder or U-net architecture as its image generator, which can be analyzed with the methods described in this article.


\subsection{Sample Application: Image to Image Translation from GOES to MRMS}

We demonstrate many of the concepts in this article for a sample application, namely estimating MRMS (radar) reflectivity from GOES satellite imagery.  Since GOES satellite imagery is available throughout the continental United States but MRMS has gaps, it would be useful to emulate MRMS in areas where it is not available.  This is a classic image to image translation task and we use a classic encoder-decoder architecture, see Fig.\ \ref{TRF_GREMLIN_figure}.
There are four input channels, namely GOES Channels 7, 9 and 13 (all infrared), and the GOES-GLM channel which indicates presence and location of lightning.
The resulting model is called GREMLIN for 
"GOES Radar Estimation via Machine Learning to
Inform NWP" \citep{hilburn2020development}.

\subsection{Receptive Fields}
\label{receptive_field_sec}

We now briefly study the increasing pattern size that the NN layers can detect.
%
\begin{figure}
    \begin{center}
    \epsfxsize=0.7\hsize \epsfbox{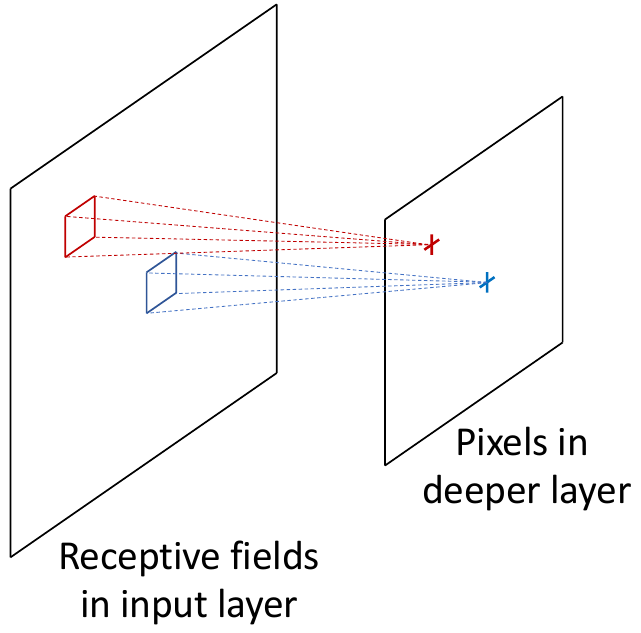}
    \end{center}
    
    \caption{Theoretical receptive field in input space shown for two pixels of a deeper layer.  The TRF is an upper bound on the spatial context used by the NN in the input image.}
    \label{TRF_two_pixels}
\end{figure}
%
%
%
\begin{figure*}
    \begin{center}
    \epsfxsize=0.95\hsize \epsfbox{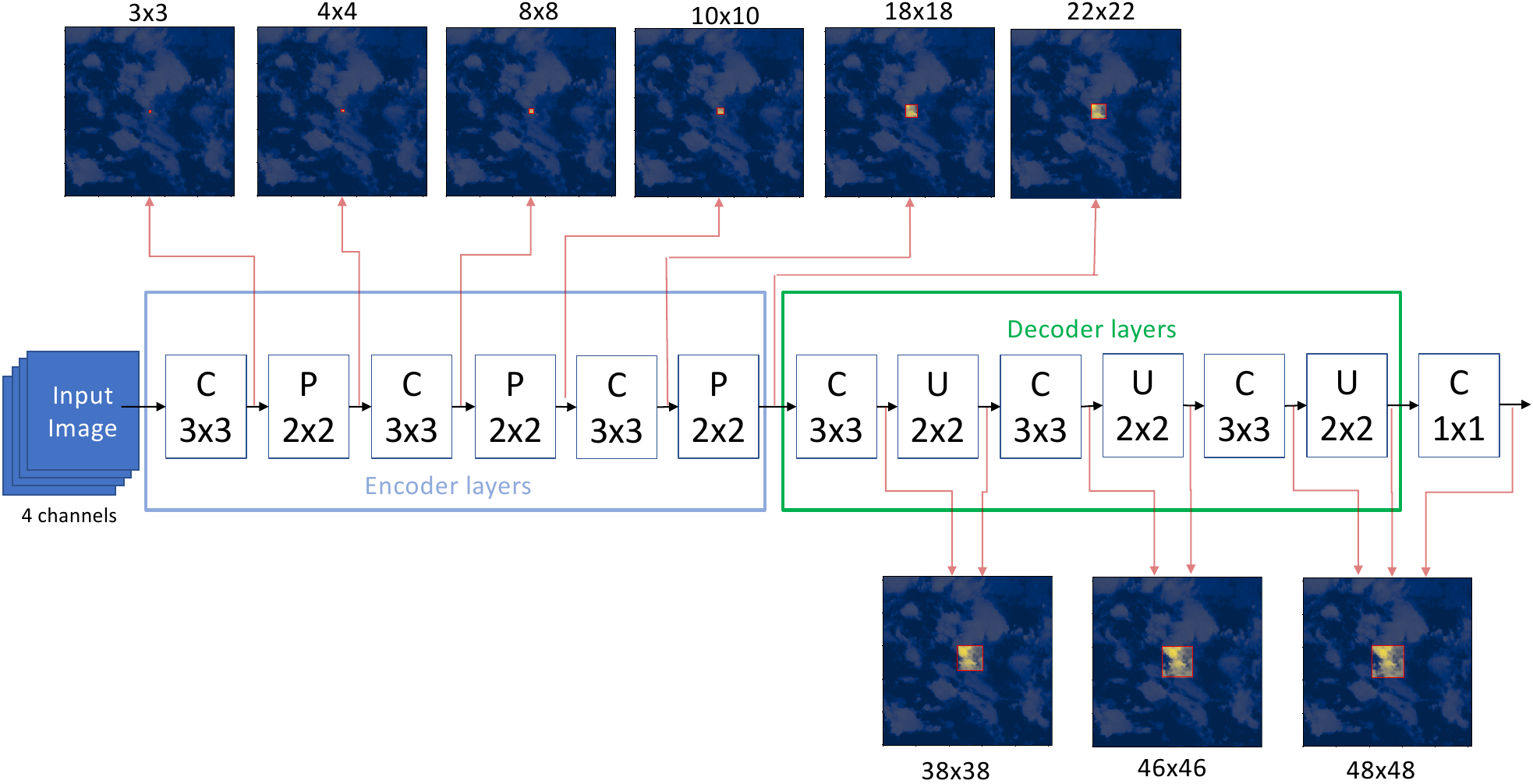}

    \end{center}
    
    \caption{Theoretical Receptive Field (TRF) of all layers of GREMLIN model visualized for Sample \#68 and input channel 13 (longwave IR). 
    The numbers on top of each image denote the TRF size in terms of pixels in the input sample.  
    The red square in each image indicates the TRF corresponding to a central pixel in the output map of a considered layer.  
    Thus the red square represents an upper bound on the spatial context in an input image that the NN can utilize in each of the layers.  The input image has 256x256 pixels and the the final layer has a TRF of 48x48 pixels.  Note that adding skip connections to the model (U-net) would not change the TRF boundaries.}
    \label{TRF_GREMLIN_figure}
\end{figure*}
%
%
The {\it theoretical receptive field (TRF)} of a NN layer is the set of all locations in the NN's input image that have a pathway to an individual output pixel of the considered layer (aka neuron in the layer's activation map).  In other words, the TRF indicates the {\it maximal size of spatial context in the input image that a considered NN layer can use}, 
see also Figure \ref{TRF_two_pixels}.
Figure \ref{TRF_GREMLIN_figure} illustrates the TRF for all layers of the GREMLIN architecture for a specific sample and for only one channel (Channel 13) of the input sample. It shows how the maximal spatial context utilized by the NN grows from layer to layer until it reaches the maximal size of 48x48 pixels. 
For a comprehensive discussion of how to calculate the TRF size for NNs, see \cite{araujo2019computing}. For a simpler explanation for many standard architectures, including all three architecture types shown in Fig.\ \ref{architecture_fig}, see \cite{le2017receptive}.

\begin{figure}
    \begin{center}
    \epsfxsize=0.90\hsize \epsfbox{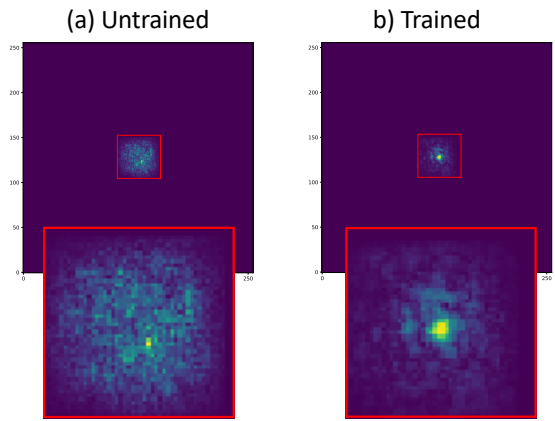}
    \end{center}
    
    \caption{Approximation of Effective Receptive Field (ERF) for final output layer of GREMLIN model (a) before training (0 epochs), i.e. using random weights, and (b) after training (100 epochs).
    This approximation is calculated with SmoothGrad \citep{smilkov2017smoothgrad} using validation sample 80 and a central output pixel, see \cite{hilburn2020development} for details. Note that the ERF for the untrained model is much more diffuse, while the trained model for this sample and location is much more focused at the center.}
    \label{ERF_evolution_training_fig}
\end{figure}
    
While the TRF determines the {\it maximal} spatial context used by each layer, \cite{luo2016understanding} show that the actual impact of pixels varies drastically within the TRF region.  They found that the impact roughly follows a Gaussian distribution, resulting in strongest impact for pixels near the TRF center and declining away from the center.  The actual distribution, known as the {\it Effective Receptive Field (ERF)}, depends on the trained NN parameters.
Figure \ref{ERF_evolution_training_fig} shows how much the ERF changes for GREMLIN during training.


TRF size can be a useful guideline when selecting how many repetitions of convolution/pooling/upsampling layers to use in NN architectures for both image translation tasks or classification tasks.
For image translation tasks the connection is obvious: 
if meteorologists can provide a rough estimate of spatial context that they think is important, then the NN architecture should be chosen to have a TRF that is at least as big, preferably already at the end of the encoder layer.  
The same reasoning applies for classification architectures, in which case one considers the TRF size at the end of the feature extraction layers.  That TRF size determines the maximal spatial context of the features used in classification.
As the ERF tends to be smaller than the TRF, the initial architecture guess is typically followed by some trial-and-error experiments with architectures of higher TRF size.
Beyond architecture design, we find calculating the TRF of the different layers (Fig.\ \ref{TRF_GREMLIN_figure}) very helpful to understand the spatial context - and thus the meteorological phenomena - the NN model is able to utilize.


\section{Using Performance Measures for NN Tuning} 
\label{performance_sec}

The meteorological community has developed many tools to evaluate the performance of algorithms used for weather and climate tasks. It seems obvious to apply such meteorological performance measures also to NN algorithms whenever they are used for the same tasks.  However that is not always happening.  The likely reason for that effect is likely the misalignment between the needs of two disciplines, computer science and meteorology. 
NN methods are usually developed by computer scientists for computer science application, which come with their own performance measures.  NN literature thus illustrate NN methods with computer science-oriented performance measures and NN software packages include them as well. 
Thus it is only natural that scientists carry over not only the methods, but also the typical performance measures that the papers teach one to use.  
{\it It is important to be conscious of this potential misalignment, i.e.\ when transferring NN methods from the computer science literature to meteorology, not to rely solely on the accompanying performance measures, and instead to also consider the full range of meteorological performance measures.}

\subsection{Examples of Underutilized Meteorological Measures}
    
\begin{figure}
    \begin{center}

    \epsfxsize=0.98\hsize 
    \epsfbox{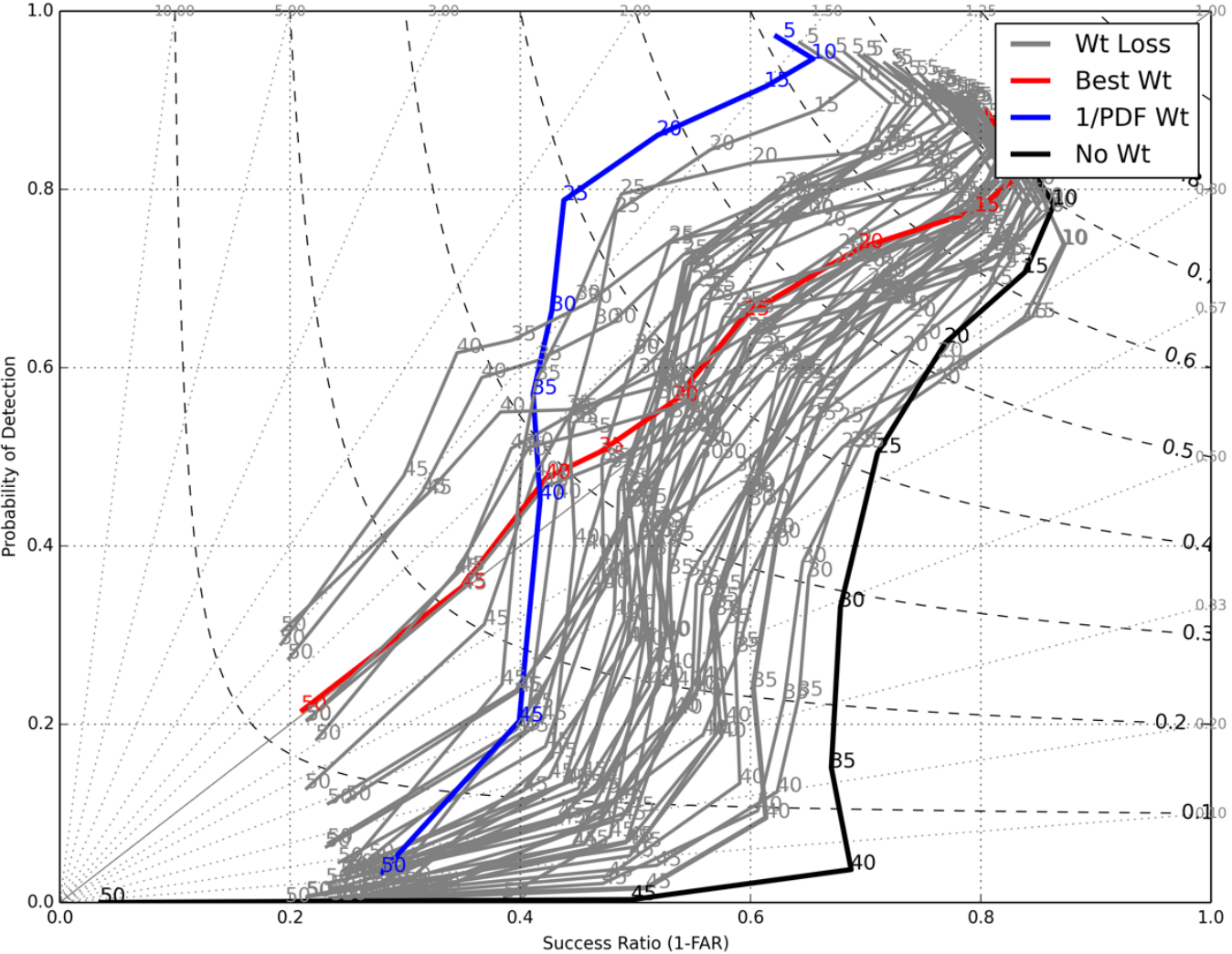}
    \end{center}
    \vspace*{-0.5cm}
    
    \caption{Use of categorical performance diagram to select weighted loss function for NN in MRMS estimation application.  
    Each line represents the performance of one NN.  Black line represents the unweighted case (plain MSE), blue line uses the inverse of value frequency as weight, and red line represents the NN selected to have the best performance.  Best performance was defined here as being closest to the diagonal line, i.e., having equal tendency to overestimate as underestimate.
    }    
    \label{categorical_performance_diagram_fig}
\end{figure}

The {\it categorical performance diagram} developed by \cite{Roebber2009-CPD} is commonly used to evaluate algorithms in meteorology, especially for weather forecasting, but we had never seen it used in the context of neural networks.
Our group recently found this diagram to be an excellent tool to tune our GREMLIN model to estimate MRMS reflectivity on a pixel-by-pixel basis. 
Evaluating the output pixel-by-pixel allows us to apply classic performance measures from meteorological forecasting problems. (In contrast, measures that evaluate the output image as a whole are discussed in the next section.)
A challenge for estimating MRMS reflectivity is that high reflectivity values are quite rare in the training samples (class imbalance), and use of a standard MSE loss function resulted in 
the NN focusing too much on providing good estimates for low reflectivity but significantly underestimated high MRMS values.  This can be corrected by adding weights to the loss function that put more emphasis on the rare (high) values. The weighted function must be chosen as a trade-off to avoid losing too much performance for lower values.
We tested the following family of weighted mean square error (MSE) functions as loss functions for an individual sample with true image, $y^{true}$, and estimated image, $y^{pred}$,
\begin{eqnarray}
    L( y^{true}, y^{pred}) &=& 
    \sum_{p=1}^P  w(y_p^{true}) \; (y_p^{pred} - y_p^{true})^2\\
    w(y_p^{true}) &=& e^{\; b \; (y_p^{true})^c},
\end{eqnarray}
where $y_p^{true}$ and $y_p^{pred}$ are the true and predicted values of output pixel $p$, $P$ is the number of pixels in the output image, and $w(y_p^{true})$ is the weight assigned to the pixel of a sample based on its true $y$-value.  Parameters $b$ and $c$ allow us to tune the weighting function, $w$.  We varied $b$ and $c$ in a grid search with $b$ ranging from 1 to 5 by increments of 0.5 and $c$ ranging from 1 to 5 by increments of 1, resulting in a total of 45 combinations. We trained a separate NN 
for each parameter combination, and plotted the performance of the resulting 45 NNs in a categorical performance diagram, see Fig.\ \ref{categorical_performance_diagram_fig}. The loss function corresponding to the red line was deemed to perform best (closest to diagonal) and chosen for our NN.  This choice greatly enhanced performance for high reflectivity values, without significant performance loss for low values.
%

A close relative of the categorical performance diagram is 
the receiver operator characteristic (ROC) curve, which can be used for NN tuning in much the same way. This has been explored in the medical literature \citep{ROC-ANN-medical,ROC-MLP-optimization}, but we have not yet seen it used in this context for meteorological applications.  

\subsection{Other Relevant Measures}

Categorical performance diagrams and ROC curves
are just two examples of a wide spectrum of meteorological performance measures to consider.  For a great overview of some interesting performance measures to consider for general machine learning algorithms in meteorology, see \cite{Gagne2015-CPD}.

Of particular interest to image translation applications are distance measures comparing pairs of images that go beyond pixel-by-pixel comparison, such as the {\it Wasserstein distance}, aka {\it Earth mover's distance}, from optimal transport theory (\citep{snow2016monge}), and the {\it Structural Similarity Index Measure (SSIM)} by \cite{wang2004image} that are already well established in remote sensing applications including in the context of NNs.
See \cite{alberga2009similarity} and \cite{tsagkatakis2019survey} for a wide variety of image evaluation measures for remote sensing.

\subsection{Loss Functions versus Auxiliary Metrics}

Performance measures can be used in neural networks in two ways, as the {\it loss function} or as {\it auxiliary metric}. The loss function tells the neural network what to optimize, so must be carefully chosen, but must also be differentiable to facilitate NN training via gradient descent methods.  
When the NN optimizes the loss function, it may do so by sacrificing other desired properties, and additional auxiliary metrics allow us to detect such trade-offs. Our recommendations are thus  
(1) to carefully choose the loss function to measure the most important (differentiable) performance property for the meteorological task, and, if needed, to define a customized loss function for that purpose rather than relying on predefined choices; and (2) to define and track one or more additional desired performance criteria as auxiliary metrics to ensure that unacceptable trade-offs are detected right away.  

One can add as many auxiliary metrics as desired in NN code, which are then evaluated after every epoch of training.  Auxiliary metrics do {\it not} have to be differentiable, which allows implementing a much bigger variety of measures than for loss functions.
For example, in our MRMS reflectivity application above we implemented the categorical performance diagram as an auxiliary metric because it is {not} differentiable. That allowed us to nevertheless determine the NN with best categorical performance out of a finite number of options. 
It also allowed us to track the evolution of categorical performance during the NN training process, which revealed that even for our tuned loss function the NN first learned predicting lower reflectivity values well, with performance for higher values improving much later in the training process.

Finally, it is common practice to include  additional terms in the loss function, combined as weighted sums. Regularization terms, which avoid NN overfitting by punishing unnecessarily large weights in the network, are a common choice. 
Another choice, which has only recently been proposed, seeks to incorporate knowledge about the underlying system into the neural network.  Namely, if there are any known constraints that the solution must obey and which are differentiable, violation of those constraints can be included in the loss function \citep{karpatne2017physics}.
For other ideas of how to incorporate knowledge about the physical system into NNs, see 
\cite{willard2020integrating}.
\cite{beucler2019enforcing} enforce physical constraints in NNs for parametrization in the NN architecture - the first example of the use of these ideas in meteorology.

\section{Interpretation through Targeted Experiments}
\label{interpretation_sec}

This section discusses ways to interpret NNs through targeted experiments.  
Section \ref{ANN_visualization_tools} describes NN visualization methods that can aid these experiments.

\subsection{Designing experiments - Generating sets of input samples to investigate}
\label{input_sets_sec}

\label{experiments_type_1_sec}

A powerful means to learn about the internals of a neural network is to split the input samples into different groups and to investigate how the NN behaves for each group.  This section provides ideas for such groupings, starting with the simplest, fairly automatic grouping and progressing to groupings that increasingly use meteorological expertise and increase in complexity and effort. In fact, the last strategy actually does not group existing input samples, but creates new sets from scratch (synthetic input samples), completely driven by meteorological expertise.

\begin{figure*}
    \begin{center}
    \epsfxsize=0.7\hsize 
    \epsfbox{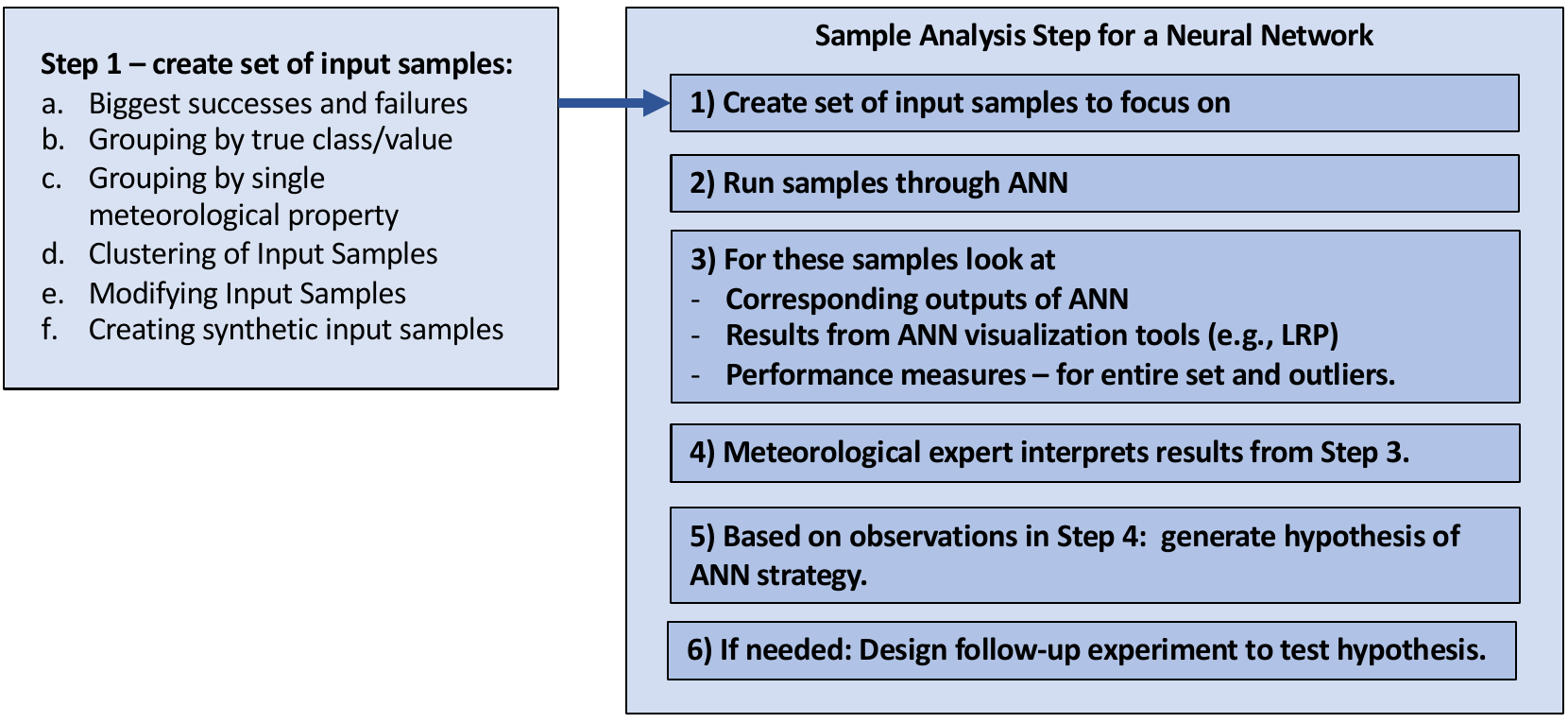}
    \end{center}
    \vspace*{-0.5cm}
    
    \caption{Sample process for analyzing a neural network architecture with focus on one set of input samples}
    \label{single_step_figure}
\end{figure*}

Once a set of input samples is chosen, for example using one of the strategies outlined in this subsection, Figure \ref{single_step_figure} shows a sample process of how it can be used to generate hypotheses.  
Note that Steps 4-6 are all ``manual'' steps to be performed by a meteorologist in conjunction with a machine learning expert. 

%


\subsubsection{Studying the biggest successes and failures}

Extreme behaviors tend to provide the biggest clues, so an analysis should usually start by studying the NN's biggest successes and failures. 
For example, to study failures for a classification task one can study several input samples that lead to false alarms, as well as several input samples that lead to misses.  Likewise, for regression tasks, one should identify and study input samples that led to gross over- or under-estimation. 
Sample questions to ask to analyze failures:
\begin{enumerate}
\item 
    Do those input samples have any obvious commonalities?  Can I find a meteorological reason why these cases would be particularly challenging?
\item
    Where in the input was the neural network looking when it made the (bad) decision?  Was it paying attention in the correct places?  Several of the ANN visualization tools can shed light on that question (see Section \ref{ANN_visualization_tools}).
\end{enumerate}
The study of successes follows analogously.
Note that out of all strategies discussed in this section, this is actually the only one that selects input samples strictly based on the NN's performance.

\subsubsection{Grouping by true class/value}

For a classification task one should always study the performance of each class separately, to see whether performance differs significantly by class.  This is of course particularly important if one of the classes represents rare events, because performance for that class might otherwise easily get overlooked.
Likewise, for a regression task one may group the input samples into groups of small, medium or high correct output values, and study how well the NN performs in each group.

\subsubsection{Grouping by single meteorological property}

It is also helpful to split the input samples by properties of the input samples that we know may greatly impact the underlying physical processes.  These might include location, time, or specific meteorological conditions.
One can then investigate the NN's dependence as follows (these are just sample criteria, not an exhaustive list):
\begin{enumerate}
\item 
    {\bf Location dependence:} How does the algorithm perform at different latitudes?  Over land versus over ocean?  Important insights can often be gained by generating a map showing NN performance by location.
\item
    {\bf Time dependence:}  Does the algorithm perform better in summer months than in winter months?  During the day or at night?
\item
    {\bf Meteorological conditions} (of course these should be entirely motivated by the application): Does the algorithm perform better for strong/weak weather events?  
\end{enumerate}
The first step above is for meteorologists to choose a categorization that they suspect might make a significant difference for the performance of the algorithm, and which is straight forward to evaluate for the input samples.  Then the input samples are assigned to the different categories (choosing value ranges for the chosen property to create discrete categories), and finally the performance of the neural network is visualized for each category. For example, maps of performance across locations often result in interesting insights that can identify information or modeling gaps, e.g., the expert may suspect that certain processes are not properly captured by the NN or that there is insufficient information in the input features for the model to be able to resolve - thus often leading to additional variables added to the input (such as latitude, time of day, etc.).

\subsubsection{Clustering of Input Samples}

Sometimes it is useful to group input samples by different meteorological regimes that are not easily described by a meteorological measure.  In that case one might apply clustering to the input samples to obtain groups representing common, well separated, meteorological scenarios, e.g., clustering images by cloud type  \citep{haynes2011major,denby2020discovering}.

\subsubsection{Modifying input samples}

In addition to selecting input samples, we can also modify them.  For example, one way to analyze the GREMLIN model for MRMS estimation is to keep the first three channels intact, but to erase all lightning in the GLM channel, to examine the effect of lightning.  Of course the type of modification, such as setting a channel to zero vs.\ assigning random values, etc., depends entirely on the application.

\subsubsection{Creating synthetic samples}
\label{synthetic_samples_sec}

\begin{figure*}
    \begin{center}
    \epsfxsize=0.7\hsize \epsfbox{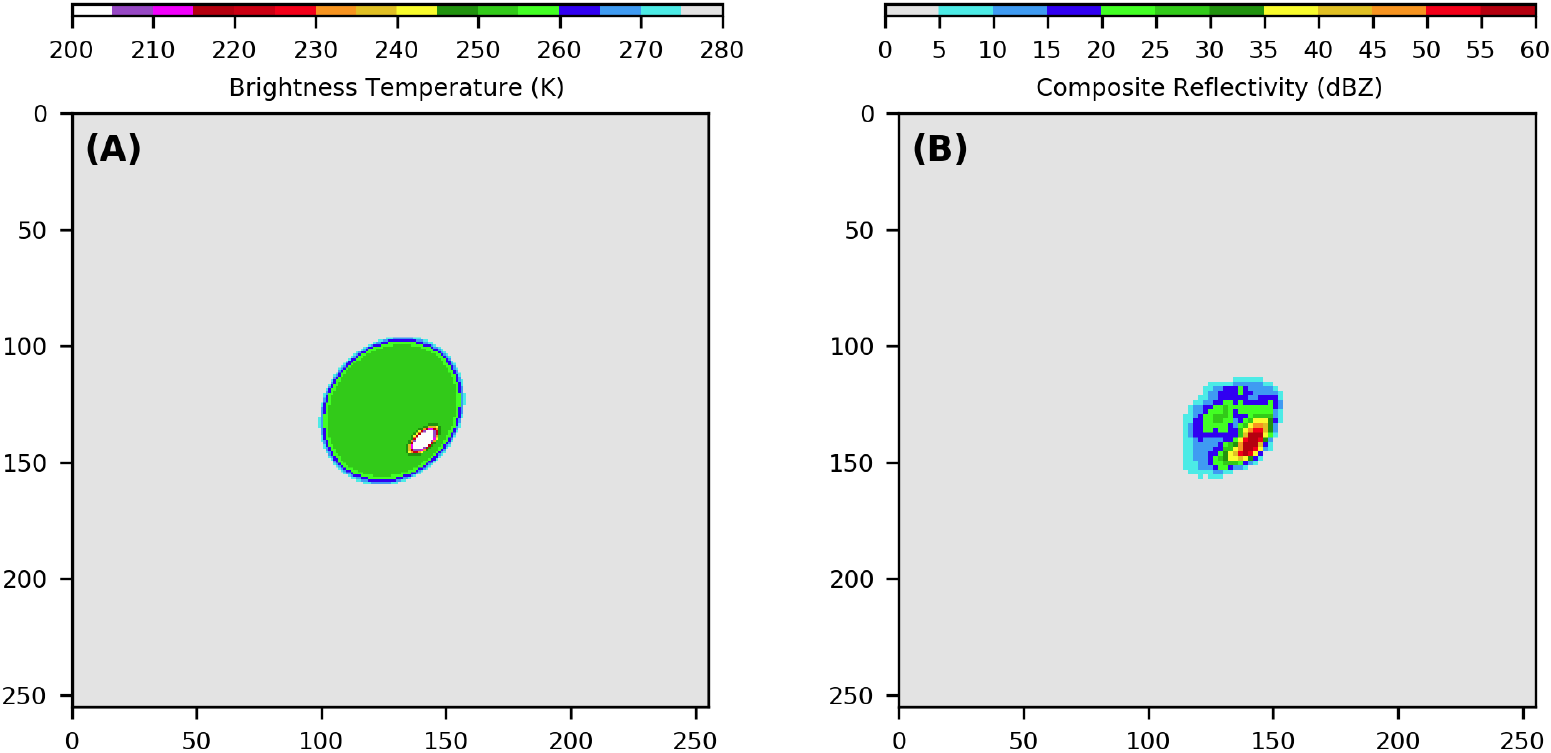}
    \end{center}
    \vspace*{-0.5cm}
    
    \caption{(a) Synthetic brightness temperature that produces maximum composite reflectivity from a grid search of over 45,000 parameter settings. 
    (b) Composite reflectivity retrieved from input in panel (a).}
    \label{synthetic_example_fig}
\end{figure*}
    
The most complex option is to create synthetic input samples that represent specific meteorological input scenarios. 
Creating such examples is challenging, and probably not possible for all applications. 
However, if achievable this provides a great tool to test the behavior of the NN by providing perfect control of the input scenario, i.e., in isolation of any other factors that might be present in an observed input sample. This can be a powerful tool in hypothesis testing, where the input scenario is specifically designed to test a single aspect. 

As example we revisit the MRMS estimation application. 
As we will see in Section \ref{ANN_visualization_tools} NN visualization methods reveal that our NN model is using radiance gradients to infer locations of high composite reflectivity. However, this is in the context of real samples that have complicated spatial patterns, which makes it difficult to answer questions such as, are sharper gradients associated with higher composite reflectivity? 
To investigate such questions, it is helpful to specify inputs with a parametric model that allows us to vary certain properties of the input in order to evaluate the sensitivity of the NN response. One such model we have had success with \citep{hilburn2020development} is a Generalized Elliptical Gaussians (GEG) model, which assumes an outer Gaussian, $G_o$, that represents the thunderstorm anvil, and an inner Gaussian, $G_i$, that represents the overshooting top. 
The synthetic brightness temperature, $T$, is a function of $(x,y)$ with the following parameters: 
location $x_0$ and $y_0$, amplitude $A$, size $S$, aspect $\alpha$, orientation $\theta$, and sharpness (exponent) $p$ for the outer and inner Gaussians, denoted with subscripts $o$ and $i$:
\begin{eqnarray}
  \hat{x}_{o,i} &=& ( x - x_{0,o,i}) \cos \theta_{o,i} - (y-y_{0,o,i}) \sin \theta_o,i ,
  \\
  \hat{y}_{o,i} &=& ( x - x_{0,o,i}) \sin \theta_{o,i}+ (y-y_{0,o,i}) \cos \theta_o,i ,
  \\
  T_{o,i} &=& \exp \left( -1 \left( \frac{\hat{x}^2_{o,i}}{2 s^2_{o,i}} + \frac{\hat{y}^2_{o,i}}{2 ( s_{o,i} \alpha_{o,i} )^2} \right)^{p_{o,i}}    \right) ,
  \\
  T &=& A_o T_o + A_i T_i.
\end{eqnarray}

Figure \ref{synthetic_example_fig} provides the synthetic input for which the NN responds with maximum composite reflectivity, based on a grid search of thousands of parameter settings. Other strongly activating patterns are similar to this one and have in common very large $p_o$ and large $p_i$, meaning that the anvil and overshooting top have very sharp $T_B$ gradients. The patterns for which the NN responds weakly have a meteorologically unphysical appearance, indicating that the NN has learned to recognize realistic overshooting top satellite signatures.

\subsection{Designing Experiments - Studying a Simpler Neural Network}

\label{experiments_type_2_sec}

\begin{figure*}
    \begin{center}
    \epsfxsize=0.95\hsize \epsfbox{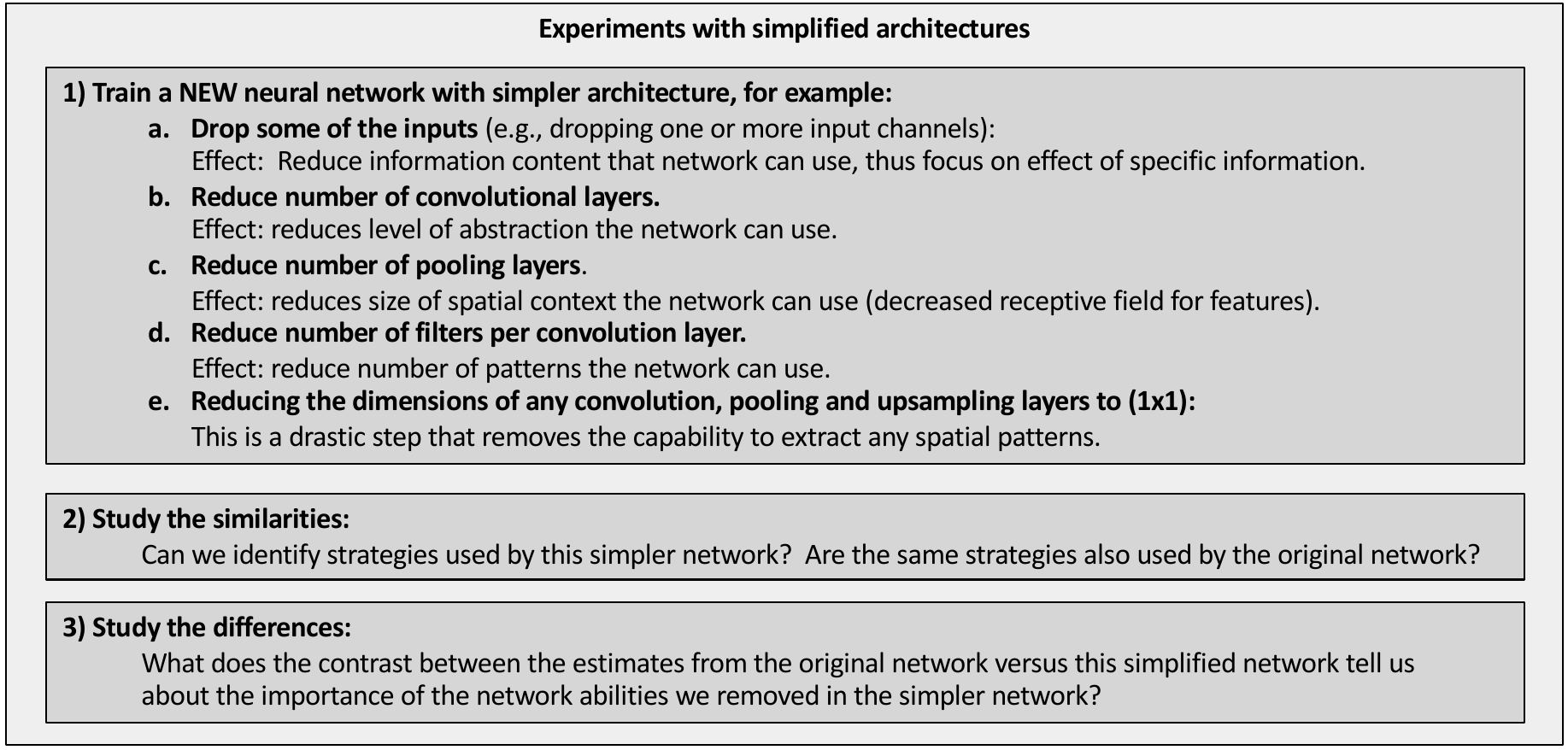}
    \end{center}
    \vspace*{-0.5cm}
    
    \caption{Studying a hierarchy of restricted NN architecture to shed light on strategies of the original NN architecture.  Key steps are to strategically define which capabilities of original model to restrict, and to 
    carefully study both similarities and differences in capabilities between the original model and restricted (simplified) models. 
    }
    \label{simpler_architecture_fig}
\end{figure*}

\begin{figure*}
    \begin{center}
    \epsfxsize=0.7\hsize \epsfbox{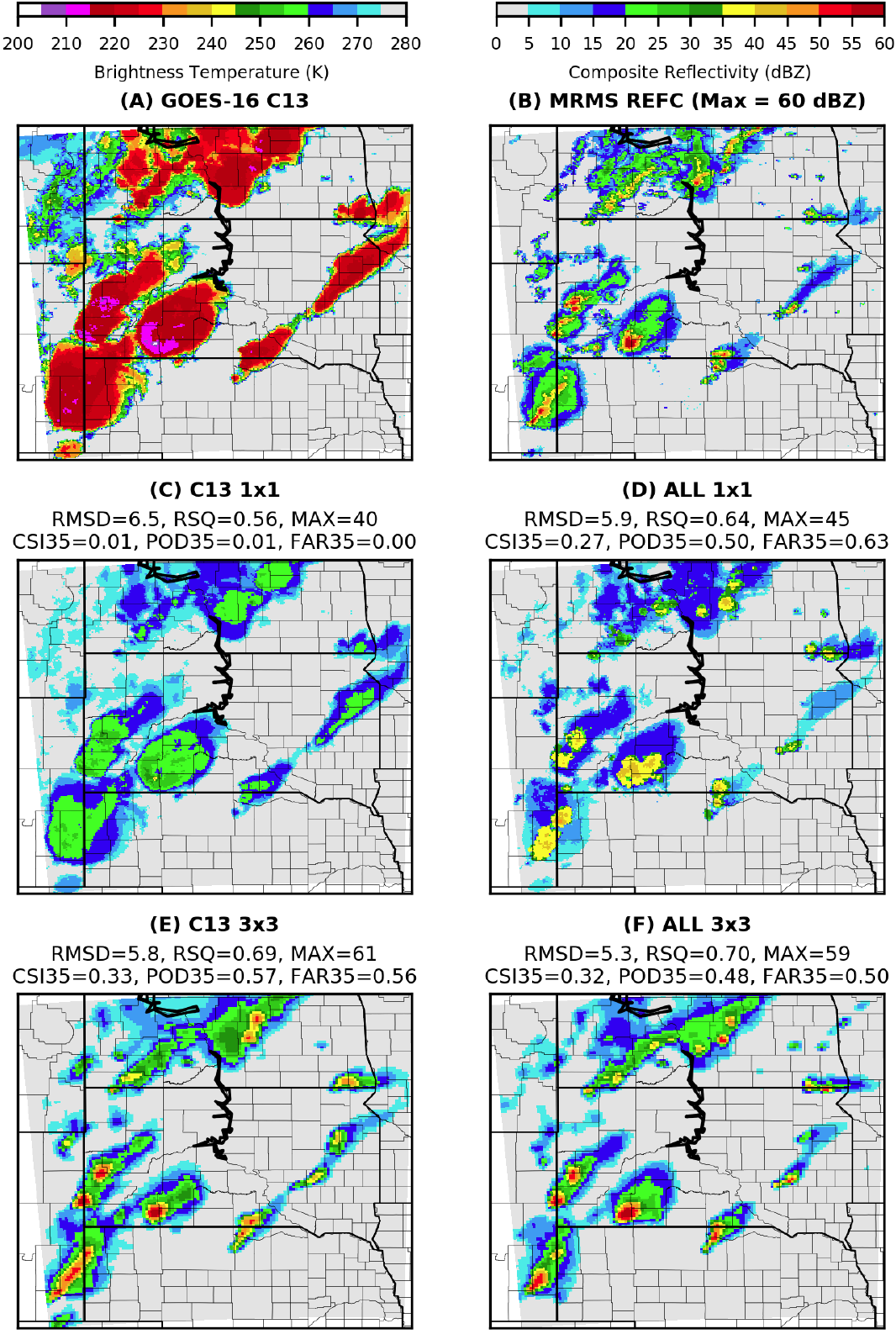}
    \end{center}
    \vspace*{-0.5cm}
    
    \caption{Example of studying a hierarchy of simpler NN architectures.  
    (a) GOES channel 13 (input); 
    (b) MRMS (true output); 
    (c-f) MRMS estimates for four different models:
    (c,d) use only (1x1) convolution and either only one input channel (c) or all input channels (d);
    (e,f) as (c,d) but using (3x3) convolution.
    Performance without spatial context (c,d) is poor. 
    Using spatial context but only one input channel (e) is much improved, with additional improvement when all channels are used (f).
    Results are for validation sample 80, 2019-07-03T22:15Z.}
    \label{simpler_architecture_example}
\end{figure*}

If studying the original network seems intractable it can be helpful to study simpler network architectures first. 
Figure \ref{simpler_architecture_fig} provides an overview of this process.  
%
%
One simplification we find particularly helpful for convolutional neural networks is to reduce the dimensions of all convolution, pooling and upsampling layers down to 1x1.  
This allows us - with only changes to a few lines of code - to turn a neural network that takes spatial context into account into a neural network that analyzes each pixel separately, eliminating any ability to use spatial context.
This allows us to study how important the spatial context is for this problem, i.e., how much performance we lose.  
A visual comparison of the qualitative difference between the NN output with or without spatial context often provides clues for the strategy used by the full network.
An example is shown in Figure \ref{simpler_architecture_example} which shows the results for four NNs of increasing complexity. As one would expect the biggest improvement increase is between Fig.\ \ref{simpler_architecture_example}(d) and \ref{simpler_architecture_example}(f), i.e.\ when we allow the model to allow spatial context.  Note how poorly the models perform without spatial context.  The methods in the next section can shine light on {\it how} specifically a NN uses spatial context.

\section{NN Visualization Methods}
\label{ANN_visualization_tools}

\cite{McGovern2019-transparent} provide an excellent overview of many neural network interpretation and visualization methods and demonstrate their use for several meteorological applications. We refer the reader to that comprehensive overview  
to learn about the methods of permutation importance, sequential (forward/backward) selection, saliency maps, gradient-weighted class-activation maps (Grad-CAM), backward optimization, and novelty detection. 
For an overview of general explainable AI methods, see \cite{XAI-book}. 
Here we focus first on the method of layer-wise relevance propagation (LRP), which is not widely known in the meteorological community (and was not studied by \cite{McGovern2019-transparent}), yet we find it to be the most promising method in the climate and weather applications we considered.  Next we contrast LRP with saliency maps, and finally we present some pitfalls and solutions for using various visualization tools for image translation versus classification tasks.

Most methods discussed below focus on analyzing a NN model by focusing on a specific sample, and providing a {\it heatmap} in the input space that can be overlayed as a mask on the input sample. 
Studying such a heatmap, in addition to the input sample and corresponding output, can provide important clues on how the NN derived the output, be that a class or an image.

\subsection{Layer-Wise Relevance Propagation}

Layer-wise Relevance Propagation (LRP) answers the question of {\it where in the input the neural network was paying the most attention} when deriving its result.  LRP and its variations were developed by \cite{Montavon2017-cl}, \cite{montavon2018methods} and \cite{Lapuschkin2019-CleverHans}.  \cite{toms2019physically} introduced 
its use for geoscience applications.
We have consistently found LRP to be the most informative NN visualization method for meteorological applications. For example, {the authors and their collaborators}
have used it to uncover strategies used by NNs 
{to discover spatial patterns of climate phenomena \citep{toms2019physically}
to detect convection in satellite imagery \citep{yoonjin2020},  
and to estimate MRMS \citep{hilburn2020development}.} 
However, to the best of our knowledge, no other research group has so far used LRP for meteorological applications.

\begin{figure*}
    \begin{center}
    \epsfxsize=0.90\hsize \epsfbox{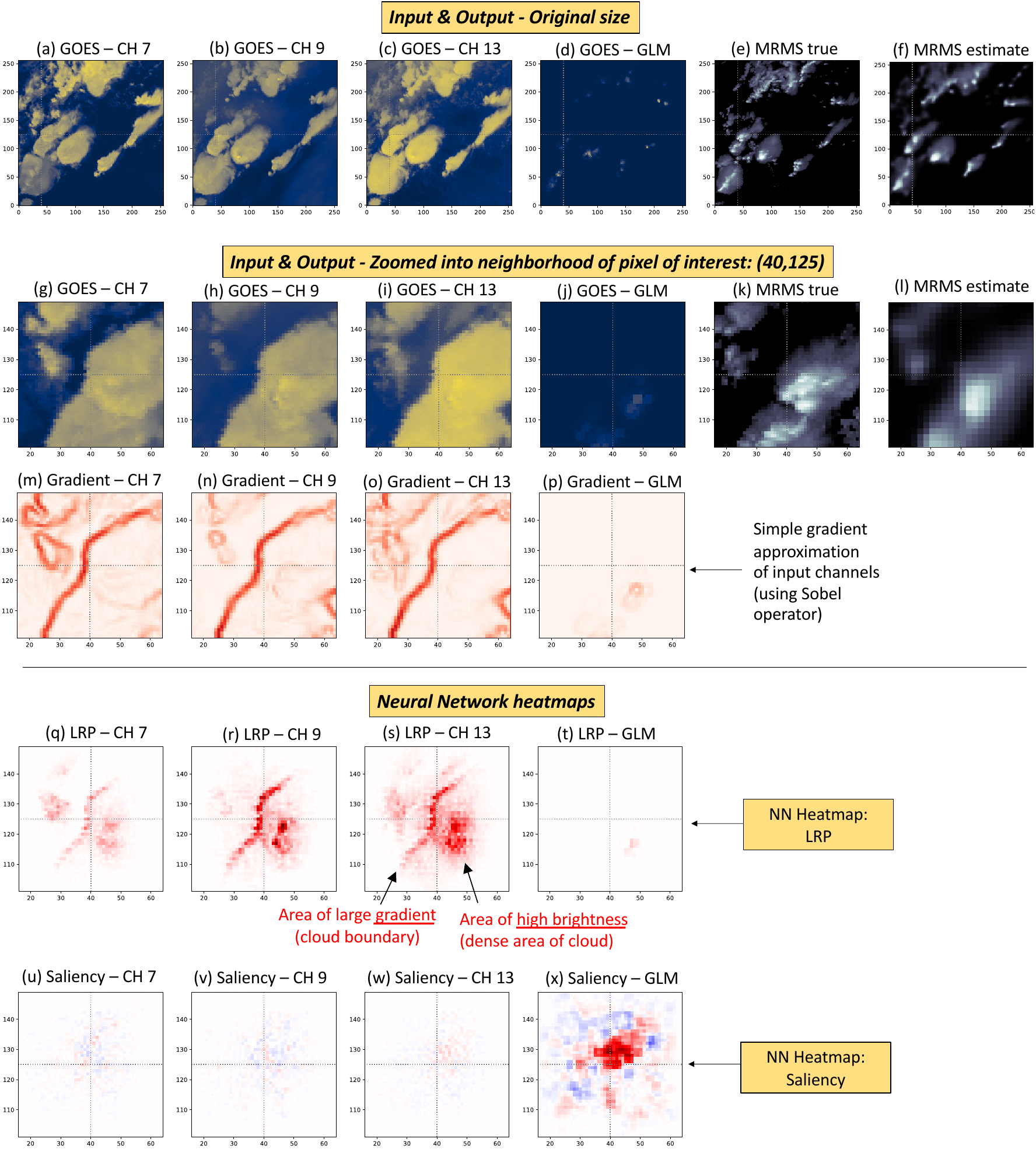}
    \end{center}
    \caption{Analysis of strategies of NN model GREMLIN for Sample 80 at output location (x=40,y=125). First row shows inputs and true/estimated outputs for full image. Second row shows the same information, but zoomed into neighborhood of desired output pixel.  (The zoom level is chosen to match the size of the TRF.)
    Third row shows the gradient of the input channels obtained by applying a Sobel operator \citep{gonzalez2002richard} to the input channels.  This auxiliary information is not used by the NN and only provided to highlight high gradient areas in the input channels for the reader. 
    Fourth row shows results for LRP heatmaps, which indicate that the NN focused here on both the gradient and high brightness regions in the three infrared channels to determine the output value.
    Fifth row shows results for saliency heatmaps, which indicate that the easiest way to increase the value of the output pixel is to increase lightning. 
    LRP and saliency maps are calculated using the iNNvestigate package \citep{alber2019innvestigate}, using the $\alpha$ - $\beta$ rule for LRP with $\alpha=1, \beta=0$.
    }
    \label{LRP_example_80}
\end{figure*}
 
Figure \ref{LRP_example_80} shows an example of LRP for GREMLIN. This example focuses on analyzing for a specific sample how GREMLIN derives the MRMS estimate for a chosen output pixel.
The first three rows provide information on the input and output for the sample. The fourth row shows the LRP heatmap, which indicates that for this sample the NN was mainly focusing on locations in the first three input channels, specifically on areas of high brightness (dense areas of clouds) and areas of large gradient (primarily cloud boundaries).  
Analyzing this and many other samples using LRP revealed the following  strategies used by our NN at any considered location: (1) Presence of strong lightning nearby takes priority, often shifting the NN's focus even in the first three input channels toward the lightning locations. (2) In the absence of lightning nearby, the NN primarily looks for areas of high cloud density or of cloud boundaries nearby. (3) If neither of the above applies, then the MRMS estimate is set to (nearly) zero.

After identifying these general strategies using LRP we then studied these strategies further using our simplified architecture experiments (Fig.\ \ref{simpler_architecture_fig},\ref{simpler_architecture_example}) to identify the importance of spatial context and of specific channels, and our synthetic experiments (Fig.\ \ref{synthetic_example_fig}) to quantify the impact of key meteorological properties in the input scenarios.

\subsection{Saliency maps}

Saliency maps, developed by \cite{simonyan2013deep}, are probably the most popular heatmap method for NN visualization.  Saliency maps answer the question {\it which input pixels should be changed to yield a maximal increase in the considered output value with minimal change?}  While this question sounds similar to the question answered by LRP, the results can differ drastically.
Consider the example in Fig.\ \ref{LRP_example_80}, where the last row shows saliency results. 
The saliency maps indicate that the most efficient way to increase GREMLIN's output value for the considered sample through modification of the input channels is to add lightning nearby.  This matches Strategy 1 discovered by our LRP analysis, namely strong sensitivity to presence of lightning.  However, saliency returns an almost identical result for any sample we considered, i.e.\ saliency was unable to identify any of the other strategies used by the NN.  
This is explained by the fact that saliency maps only look at {\it local gradients} in the input space, which can be limiting, as demonstrated by \cite{Montavon2017-cl}, while LRP takes a more global view.
On the other hand, saliency is widely available for any network type, while LRP implementations are currently available only for a small set of NN architectures.

\subsection{Visualization methods for image translation versus classification}

Finally we discuss some considerations and pitfalls encountered when using heatmap methods for image translation versus classification tasks.
Heatmaps were developed for classification tasks, so some caution must be applied when using them for other tasks.  First of all, to generate NN heatmaps one must specify an input sample and an output neuron to be analyzed. In the case of classification architectures the output neuron corresponds to a specific class.  
If no class is prescribed the algorithm uses as default the class assigned to the sample by the NN, which is typically a meaningful choice.  
In contrast, in an image translation task an output neuron corresponds to a specific pixel which the user must specify for the analysis to be meaningful.  If not specified, the analysis typically progresses without warning, but the results are not meaningful, because the NN now misinterprets the values of the output image as class probabilities rather than different pixels.

Secondly, gradient-weighted class-activation maps (Grad-CAM), a method used with great success by \cite{McGovern2019-transparent}  for classification tasks, requires the NN to have at least one fully connected layer.  Thus Grad-CAM is not applicable to typical NN architectures for image translation. Note that implementations typically do not check for this, and might return heatmaps even if the method does not apply. It is thus the responsibility of the user to ensure that an applied method is indeed applicable.

Thirdly, we tried to apply the method of optimal input, aka feature visualization \citep{Olah2017-jv,Olah2018-uf}, 
to our MRMS prediction task, but it always yielded adversarial examples rather than physically meaningful results.  We suspect that optimal input is better suited for classification tasks, see for example their successful use by \cite{McGovern2019-transparent}, because different classes might have dominant patterns that act as attractors for the optimal input, and no equivalent patterns might exist in typical image translation tasks.  However, this is a topic of active investigation.


\section{Conclusions}
\label{conclusions_sec}

This article highlights many strategies and practical considerations for the development and interpretation of neural networks for meteorological applications, including the concept of effective receptive fields, discussion of performance measures, and a long list of potential methods for NN interpretation, with synthetic experiments and LRP heatmaps emerging as particularly useful tools.
A common thread that emerged throughout is that NN interpretation needs to be driven by meteorological experts via specific questions or hypotheses to be investigated for the NN. 
Meteorologists have a crucial role to play to develop creative, meteorologically motivated standards for all aspects of NN development and interpretation, and such efforts will go a long way to create trustworthy neural networks for operational use in meteorology.

\vspace*{0.5cm}

\acknowledgments

This work was supported in part by the National Science Foundation through grant HDR-1934668 (Ebert-Uphoff) and by NOAA's GOES-R Program through award NA19OAR4320073. We would like to thank NOAA RDHPCS for access to the Fine Grain Architecture System on Hera, without which this research would not have been possible.
We are grateful to the developers of the iNNvestigate package \citep{alber2019innvestigate} available at 
\url{https://github.com/albermax/innvestigate}.


 \bibliographystyle{ametsoc2014}
 \bibliography{references}

\end{document}